\documentclass[aps,twocolumn,nobibnotes,prl]{revtex4}
\usepackage{amsmath,epsfig,graphicx,amssymb}
\usepackage{bm}%
\expandafter\ifx\csname package@font\endcsname\relax\else
 \expandafter\expandafter
 \expandafter\usepackage
 \expandafter\expandafter
 \expandafter{\csname package@font\endcsname}%
\fi

\begin{document}
\title{ADIABATIC TRANSPORT OF BOSE-EINSTEIN CONDENSATE IN DOUBLE- AND
TRIPLE-WELL TRAPS}
\author{V.O. Nesterenko$^1$, A.N. Novikov$^{1,2}$, F.F. de Souza Cruz$^3$,
E.L. Lapolli$^3$}
\affiliation{$^{1}$
Bogoliubov Laboratory of Theoretical Physics,
Joint Institute for Nuclear Research, Dubna, Moscow region, 141980,
Russia}
\affiliation{$^{2}$
Tver State University, Tver, Russia}
\affiliation{$^{3}$ Departamento de
Fisica, Universidade Federal de Santa Catarina, Florianopolis, SC,
88040-900, Brasil}
\date{\today}

\begin{abstract}
 By using a close similarity between multi-photon and tunneling population
transfer schemes, we propose robust adiabatic methods for the transport
of Bose-Einstein condensate (BEC) in double- and triple-well traps. The
calculations within the mean-field approximation (Gross-Pitaevskii equation)
show that irreversible and complete transport takes place even in the presence
of the non-linear effects caused by interaction between BEC atoms. The transfer
is driven by adiabatic time-dependent monitoring the barriers and well depths.
The proposed methods are universal and can be applied to a variety of
systems and scenarios.
\end{abstract}

\maketitle

The trapped Bose-Einstein condensate (BEC) is now widely recognized
as a source of new fascinating physics and interesting cross-over
with other areas \cite{BEC_reviews,crossover}.
In particular, a large attention is paid to dynamics of weakly
bound condensates. Here two main systems are usually
considered: multi-level BEC  in a single potential well (see
\cite{Nista_08} and references therein) and a single
BEC in a multi-well trap \cite{Smerzi_PRL_97,Raghavan_PRA_99,
Zhang-Mossmann,Graefe_PRA_06,Pitaevskii_UFN_06,exper}.
In the former case, BEC components are formed by atoms in different
hyperfine levels coupled by the resonant laser irradiation.
In the second case, all BEC atoms are in the same state but the trap
is separated by barriers into a set of potential wells and BEC dynamics
reads as tunneling between the wells.
This case is realized in multi-well traps \cite{exper}, arrays of
selectively addressable traps \cite{arrays}
and optical lattices \cite{Morsch_RMP_06,Pitaevskii_UFN_06}.

The multi-component and multi-well systems
are quite similar and can be treated on the same footing as a
BEC with (weakly) coupled fractions. In the former,
the fractions coincide with BEC components while in the later
with populations of the wells. Even for the  constant
coupling, dynamics of multi-fractional  BEC is very reach and varies
from different kinds of Josephson oscillations (in analogy with
internal and external Josephson effects in weakly bound superconductors,
see discussion in \cite{Raghavan_PRA_99}) to Mott insulator transitions
and other transfer regimes pertinent to electrons in a crystal lattice
\cite{Morsch_RMP_06}. The dynamics strictly depends on the interplay
between the interaction of BEC atoms and the coupling
\cite{Smerzi_PRL_97,Morsch_RMP_06,Pitaevskii_UFN_06}.

For a constant coupling, BEC dynamics is mainly reduced to
oscillating fluxes of atoms between the fractions. That
was a subject of intense investigation during the last decade
\cite{Nista_08,Smerzi_PRL_97,
Raghavan_PRA_99,Zhang-Mossmann,Graefe_PRA_06,
Pitaevskii_UFN_06,exper}.
At the same time, still a little was done for exploration
of {\it irreversible} transport of BEC between its fractions  when
their populations do not oscillate but evolve by irreversible
and controlled way. Specifically, BEC transport means that condensate,
being initially in one of the fractions, is then completely transferred
to another (target) fraction and kept there.  Being realized, BEC
transport could open intriguing perspectives for investigation of
topological states \cite{topol_1,topol_2}, generation and exploration of
geometric phases (which may serve as promising information carriers
\cite{quantum_info}), various implementations in quantum computing (e.g.
to build analogs of Josephson qubits in superconductors \cite{JQ}), etc.

Nowadays adiabatic population transfer methods seem to be most promising
to carry out the transport of BEC. A variety of suitable adiabatic methods
was already developed for the aims of atomic and molecular spectroscopy
\cite{Vitanov}, between them Stark Chirped Rapid Adiabatic Passage (SCRAP)
\cite{Yatsenko_PRA_99} and Stimulated Rapid Adiabatic Passage (STIRAP)
\cite{Berg_STIRAP}. These methods exploit two-photon population transfer
schemes and intend to excite non-dipole states in electronic systems.
However, the methods are quite general and can be upgraded to
other systems and scenarios, see e.g.
\cite{Ne_clusters,Ne_Bars_08,Eckert_PRA_04}.

The aim of the present paper is to adapt the SCRAP and STIRAP ideas
for BEC transport and develop the relevant population transfer schemes.
For this aim the proper time-dependent protocols of the system
parameters (well depths and barrier penetrabilities) will be proposed.
We will concentrate on double- and triple-well traps where
the necessary control of the parameters can be easily realized, e.g.
the barriers can be monitored by varying
depths and separations of the wells \cite{exper,arrays}.

By our knowledge, the SCRAP has not still employed in BEC transport.
Quite recently an alternative technique, Rabi switch,
was suggested, where time-dependent Rabi coupling of a fixed duration
was used to transfer solitons and vortices \cite{Nista_08}.
Here we propose in principle another scheme where both Rabi frequency and
detuning are adiabatically controlled. Such SCRAP-like scheme has the
advantage to be rather insensitive to the process parameters, e.g.
the coupling duration \cite{Vitanov,Yatsenko_PRA_99}.

As for STIRAP, it was  already proposed to transfer individual atoms
\cite{Eckert_PRA_04} and BEC \cite{Graefe_PRA_06}. In BEC
the nonlinearity due to the interatomic interaction $U$ was shown to
be detrimental for the adiabatic transfer (like dynamical
Stark shifts in multi-level system) \cite{Graefe_PRA_06}. As a remedy,
a large detuning $\Delta$  of the well depths (trap asymmetry) was
proposed. We will show that, unlike conclusions \cite{Graefe_PRA_06},
both large non-linearity and asymmetry are detrimental for STIRAP but
the transfer is robust at their modest values. Instead, in SCRAP-like
scheme a large time-dependent detuning $\Delta (t)$ is useful and even
crucial.

Altogether, we will develop and justify simple and effective adiabatic
schemes for BEC transport in double-, triple- and multi-well traps.

Our calculation have been performed in the mean-field approximation
by using the non-linear Schr$\ddot o$dinger, or Gross-Pitaevskii
equation \cite{GPE} for BEC:
\begin{equation}
\label{eq:NLSE}
i\hbar{\dot \Psi}({\vec r},t) = [-\frac{\hbar^2}{2m}\nabla^2
+ V_{ext}({\vec r},t)
+ g_0|\Psi({\vec r},t)|^2]\Psi({\vec r},t)
\end{equation}
where the dot means time derivative, $\Psi({\vec r},t)$ is the classical
order parameter of the system, $V_{ext}({\vec r},t)$ is the external trap
potential involving both (generally time-dependent) confinement and coupling,
$g_0=4\pi a/m$ is the parameter of interaction
between BEC atoms, $a$ is the scattering length and $m$ is the atomic
mass.

For BEC with M fractions, the order parameter can be expanded as
\cite{Raghavan_PRA_99}
\begin{equation}\label{eq:Psi}
\Psi({\vec r},t)=\sqrt{N}\sum_{k=1}^M \psi_k(t)\Phi_k({\vec r})
\end{equation}
where $\Phi_k({\vec r})$ is the static ground state solution of (\ref{eq:NLSE})
for the isolated (without coupling) k-th well
\cite{note} and $\psi_k(t)=\sqrt{N_k(t)}e^{i\phi_k(t)}$ is the amplitude
related with the relative population of the k-th well $N_k(t)$
and corresponding phase $\phi_k(t)$. The total number of
atoms $N$ is fixed:
$\int d\vec r |\Psi({\vec r},t)|^2/N=\sum_{k=1}^M N_k(t) = 1 \; .$

Being mainly interested in evolution of populations $N_k(t)$,
we dispose by integration of the spatial distributions
$\Phi_k({\vec r})$  and finally get
\cite{Smerzi_PRL_97,Raghavan_PRA_99}
\begin{equation}\label{psi(t)}
  i{\dot \psi}_k = [E_k(t)+ UN|\psi_k|^2]\psi_k
- \sum_{j \ne k}^M \Omega_{kj}(t) \psi_j \;
\end{equation}
where
\begin{equation}\label{Om}
 \hbar \Omega_{kj} (t) = -
 [\frac{\hbar^2}{2m}\nabla\Phi^*_k \cdot\nabla\Phi_j
 +\Phi^*_k V_{ext}(t)\Phi_j]
\end{equation}
is the coupling between BEC fractions,
\begin{equation}\label{E}
  \hbar E_k(t)= \int d{\vec r} \;
  [\frac{\hbar^2}{2m}|\nabla\Phi^*_k|^2
  +\Phi^*_k V_{ext}(t)\Phi_k]
\end{equation}
is the potential depth, and
\begin{equation}\label{U}
  \hbar U= g_0\int d{\vec r} \; |\Phi_k|^4 \;
\end{equation}
labels the interaction between BEC atoms. The values $\Omega_{kj} (t)$,
$E_k(t)$, and $U$ have frequency dimension.

We use the coupling of the Gauss form with a common amplitude $K$:
\begin{equation}\label{Omega}
\Omega_{kj}(t)=K {\bar\Omega}_{kj}(t), \quad
\bar{\Omega}_{kj}(t)=\exp\{-\frac{(t_{kj}-t)^2}{\Gamma_{kj}^2}\}
\end{equation}
where $t_{kj}$ and $\Gamma_{kj}$ are  centroid and
width parameters. Then, dividing (\ref{psi(t)}) by $1/2K$, one gets
\begin{equation}\label{spsi(t)}
  i\hbar \dot{\psi}_k = [\bar{E}_k(t)
  +\Lambda |\psi_k|^2]\psi_k
- \frac{1}{2}\sum_{j \ne k}^M \bar{\Omega}_{kj}(t) \psi_j
\end{equation}
where
\begin{equation}\label{Lambda}
\bar{E}_k(t)=E_k(t)/2K, \quad \Lambda=UN/2K
\end{equation}
and the time is scaled as $2Kt \to t$ and so is dimensionless.
In (\ref{spsi(t)}) the key
parameter $\Lambda$ is responsible for the interplay between the
coupling and interaction. In principle, one may further upgrade (\ref{spsi(t)})
by canonical transformation of $N_k$ and $\phi_k$ to other unknowns,
population imbalancies and phase differences, and thus removing
from (\ref{spsi(t)}) the integral of motion $N$ \cite{Ne_BEC}.
It should be emphasized that, unlike the previous studies  of the
{\it oscillating} BEC fluxes in traps with {\it constant} parameters
\cite{Nista_08,Smerzi_PRL_97,Raghavan_PRA_99,Zhang-Mossmann},
we will deal with irreversible BEC {\it transport}
by monitoring {\it time-dependent}
parameters, namely depths $E_k(t)$ and couplings
$\Omega_{kj}(t)$.

\begin{figure}
\includegraphics[height=6.3cm,width=5.7cm,angle=-90]{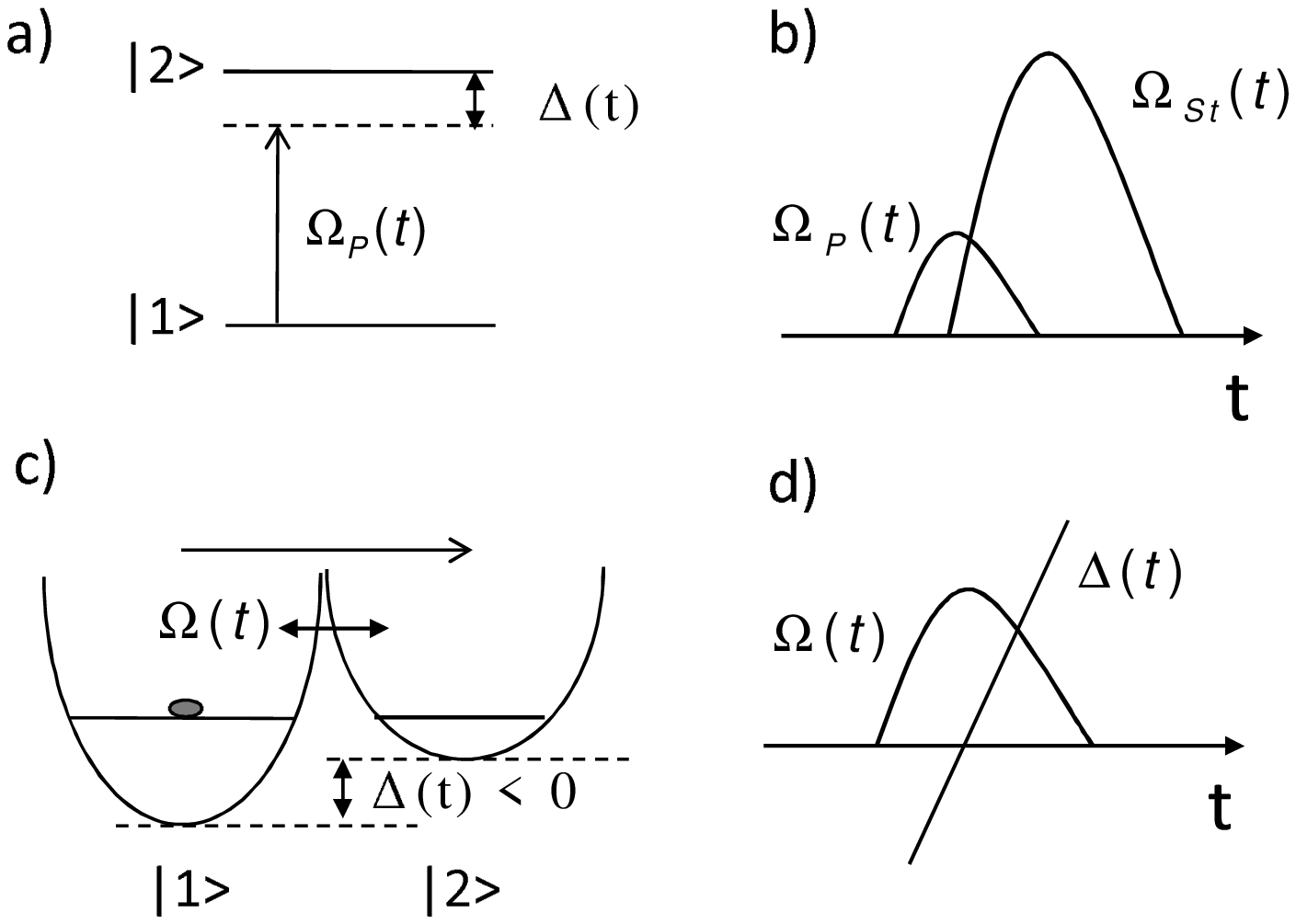}
\vspace{0.2cm}
\caption{\label{fig:fig1_2lev}
Schemes of the adiabatic population transfer in double-level
a)-b) and double-well c)-d) systems.
a) Double-level scheme with off-resonant pump laser pulse
$\Omega_P(t)$ and detuning $\Delta (t)$.
b) Time evolution of pump $\Omega_P(t)$ and Stark $\Omega_{St}(t)$
   laser pulses.
c) Double-well scheme with barrier coupling $\Omega (t)$ and
   detuning $\Delta (t)$. The condensate (dot at the left well)
   should be transferred to the right well, as indicated by the
   long arrow.
d) Time evolution of the coupling and detuning.
}
\end{figure}

In the present study, the cases with M=2 and 3 are explored. For M=2 the SCRAP
scheme for a double-level system \cite{Vitanov,Yatsenko_PRA_99} and its
counterpart proposed for BEC transport in a double-well system are illustrated
in Fig. 1, see plots a)-b) and c)-d), respectively. In both cases, we deal with
time-dependent detuning and coupling. In the double-level system, the detuning
$\Delta(t)$ (= dynamical Stark shift) and the coupling are generated by the
Stark $\Omega_{St}(t)$ and pump $\Omega_{P}(t)$ pulses, respectively. Instead,
in the double-well system the detuning $\Delta(t)=\bar{E}_2(t)-\bar{E}_1(t)$ is
produced by variation of well depths $E_k(t)$, while the coupling $\Omega(t)$
is determined by tunneling (\ref{Om}) between the barriers. The later can be
easily monitored by variation of the well separation. Obviously, the cases a)
and c) are very similar: they involve two states $|1\rangle $ and $|2\rangle $
together with the time-dependent coupling and detuning. And in both cases the
adiabatic population transfer takes place when the detuning and coupling are
simultaneously and slowly varied. To make the transfer irreversible, the system
has to cross only once the resonance (symmetry) point. For this reason, the
pump pulse should cover only one of the flanks (left in the plot b)) of the
Stark pulse. The same aim is achieved in the double-well system  by crossing
once the symmetry point $\Delta(t)=0$ during the detuning rise, see plot d). In
both two-level and two-well cases the adiabatic condition \cite{Vitanov}
\begin{equation}\label{eq:adia_two_wells}
|\dot{\theta}(t)|\ll \sqrt{\Omega^2_P(t)+\Delta^2(t)}
\end{equation}
has to be fulfilled, where $\theta$ is the mixing angle of the states
$|1\rangle $ and $|2\rangle $ during the adiabatic evolution.

\begin{figure}
\includegraphics[height=7.8cm,width=5.7cm,angle=-90]{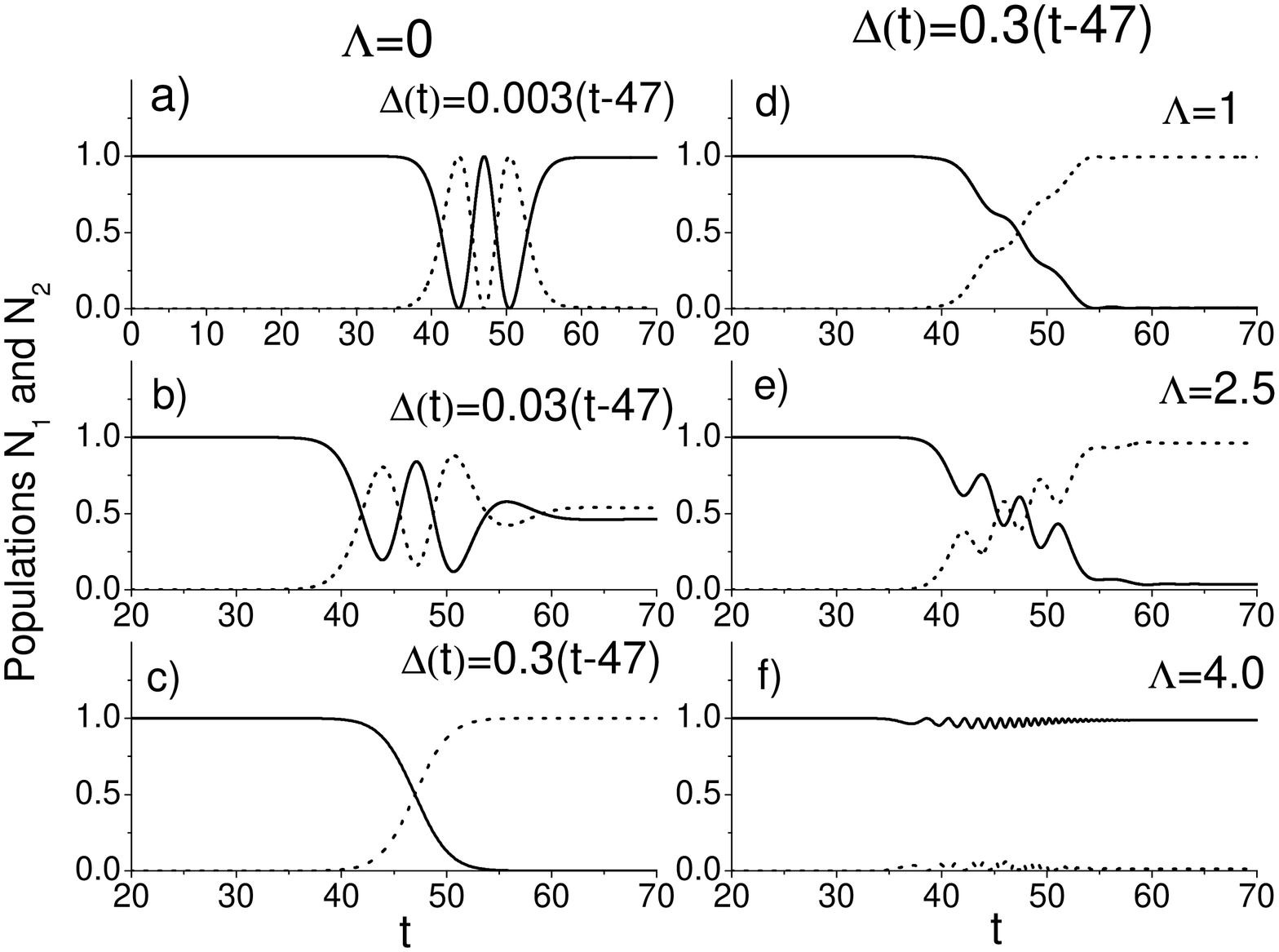}
\vspace{0.2cm}
\caption{\label{fig:fig2_2pop}
Populations $N_1(t)$ (solid curve) and $N_2(t)$ (dotted curve) calculated
with different interaction $\Lambda$ and detuning $\Delta (t)$.
Left panels: no interaction ($\Lambda =0$) and different rates of
detuning $\Delta (t)$. Right panels:  the fixed rate of detuning,
$\Delta (t)=0.3(t-47)$, and different interaction.
The coupling $\bar{\Omega}_{12}(t)$ is characterized by
$t_{12}$=47 and $\Gamma_{12}$=7.1.
Time is dimensionless. The initial phase difference is zero.
}
\end{figure}

In Fig. 2 the results of our calculations for the double-well case
are demonstrated. The plots a)-c) show that the population transfer
improves with increasing the detuning rate $\dot{\Delta}(t)$ from
0.003 to 0.3. For $\dot{\Delta}(t)$=0.3 the transfer is
complete. So, the robust transport of BEC needs a rapid change of
the detuning, which agrees with the adiabatic condition.
(\ref{eq:adia_two_wells}).
Indeed, the rapid rise of
$\Delta(t)$ makes $\sqrt{\Omega^2(t)+\Delta^2(t)}$ large
enough not only at the crossing point $\Delta(t)=0$ where $\Omega(t)$
is maximal but also at the remote flanks of $\Omega(t)$. As a result,
the condition (\ref{eq:adia_two_wells})
takes place for a long time, which favors the adiabatic transfer. Further,
as is seen from the plots d)-e), our scheme provides the complete
transport even for the interacting condensate ($0 \le \Lambda \le 2.5$).
The transfer is fine at $\Lambda = 1.0$ and still persists, though
with fluctuations, at $\Lambda = 2.5$. For stronger interaction
$\Lambda = 4.0$ the transfer breaks down. So, provided the rapid detuning
change, our scheme ensures a robust and complete transport of the
BEC even under a modest interaction.

\begin{figure}
\includegraphics[height=7.0cm,width=5.0cm,angle=-90]{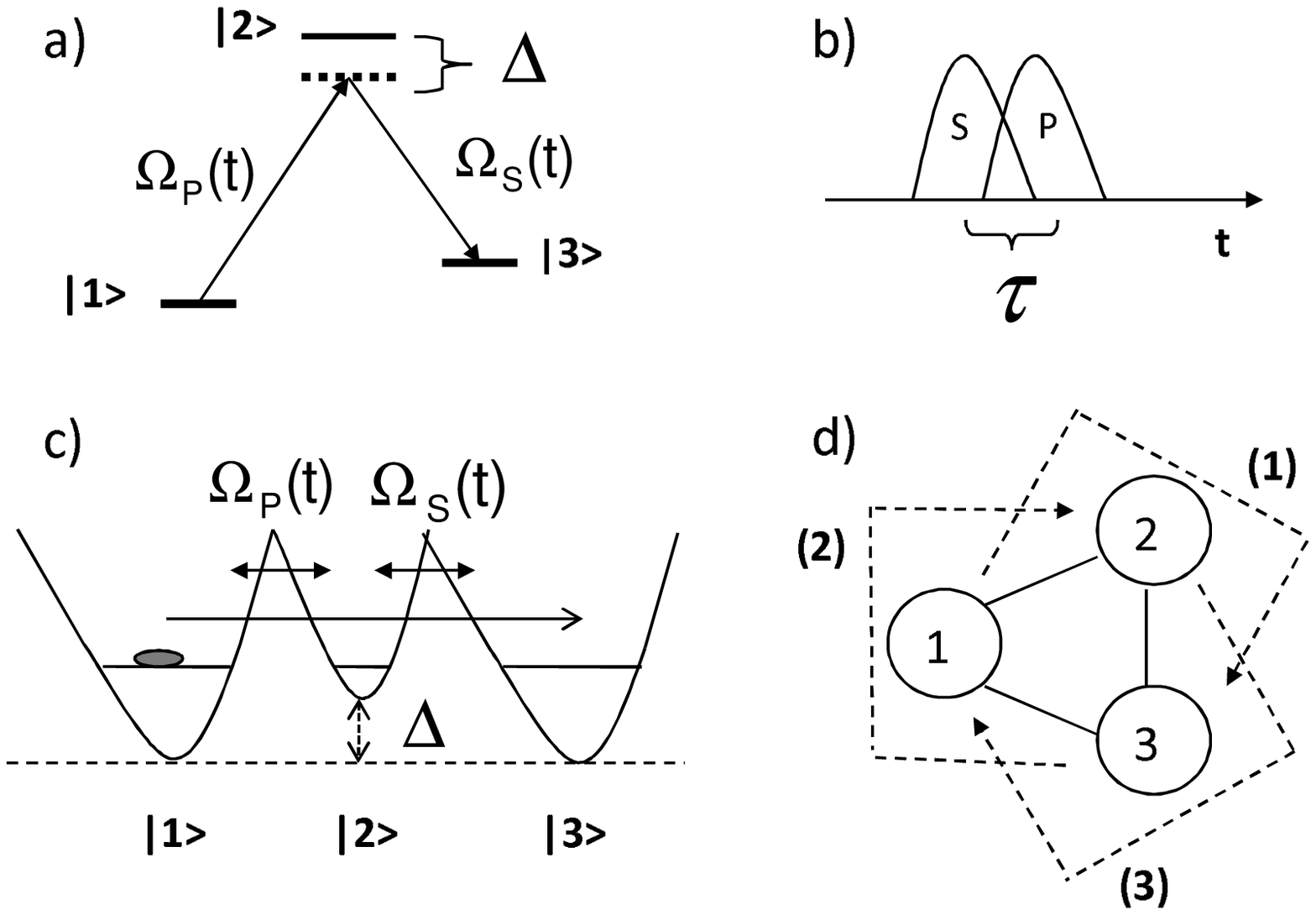}
\caption{\label{fig:fig4_3lev}
STIRAP schemes in three-level and triple-well systems.
a) Three-level scheme with pump $\Omega_P(t)$ and Stokes $\Omega_S(t)$
laser pulses and detuning $\Delta$.
b) Counterintuitive sequence of pump and Stocks
pulses with the overlap $\tau$.
c) Triple-well scheme with the barrier couplings $\Omega_P(t)$
   and $\Omega_S(t)$ and detuning $\Delta$.
d) Three-step STIRAP transfer in the circular well chain. The dash
arrows mark three STIRAP steps.
}
\end{figure}

In Fig. 3 the STIRAP population transfer scheme is illustrated for triple-level
a) and triple-well c) systems. Now we deal with three states ($|1\rangle$,
$|2\rangle$ and $|3\rangle$), pump and Stocks time-dependent couplings
($\Omega_P(t)$ and $\Omega_{S}(t)$), and constant detuning $\Delta$. The
detuning is not crucial here and can be omitted \cite{Vitanov,Berg_STIRAP}.
Instead, the counterintuitive sequence of the couplings (Stocks precedes pump)
and their overlap during sufficiently long time $\tau$ are important, see plot
b). Only under such conditions it is possible to provide a slow adiabatic
evolution from the initial state $|1\rangle$ to the target state $|3\rangle$.
The adiabatic condition $|\dot{\theta}(t)|\ll \Omega (t)$
\cite{Vitanov,Berg_STIRAP} can be written as
\begin{equation}
\frac{\Omega_S(t)\dot{\Omega}_P(t)-\Omega_P(t)\dot{\Omega}_S(t)}
{(\Omega_S^2(t)+\Omega_P^2(t))^{3/2}}\ll 1
\; \mbox{or} \; \Omega_{\tau} \gg \frac{d}{\sqrt{2}\; \Gamma^2}
\end{equation}
where $\Omega_{\tau}$ is the average amplitude of the pump and Stokes couplings
during the overlap time $\tau$, $d=t_{12}-t_{23}$ is the relative pump-Stokes
shift, and $\Gamma$ is the width parameter in (\ref{Omega}). Following STIRAP
practice, one may take $\Omega_{\tau} \approx 0.5K$ and $\tau \approx \Gamma$.
Then it is seen that the adiabatic transfer needs
a strong coupling $K$ and/or a long overlap time $\tau$.

If the couplings $|1\rangle\leftrightarrow |2\rangle$ and
$|2\rangle\leftrightarrow|3\rangle$ are supplemented by the
additional one $|3\rangle\leftrightarrow|1\rangle$, then
one may launch, via three STIRAP steps,
$|1\rangle \to |3\rangle, \; |3\rangle \to |2\rangle, \; |2\rangle \to |1\rangle$,
a cyclic adiabatic transfer, see plot 3d). Such a cyclic adiabatic
transport can be used to generate the geometric Berry phase \cite{Ne_BEC}.

In Fig. 4, results of our calculations for 3 sequential STIRAP-like
transfers (constituting altogether the cyclic evolution) are considered.
In the plot a) we see the robust STIRAP transfer at $\Lambda=0$, i.e.
without the interaction between BEC atoms and related noninearity.
Being initially in the first well ($N_1=1, N_2=N_3=0$), BEC undergoes
the sequence of full population transfers,
$|1\rangle \to |3\rangle, \; |3\rangle \to |2\rangle, \; |2\rangle \to |1\rangle$,
 and finally fully returns back to the well $|1\rangle$.
The transfer survives at modest interaction, $\Lambda \le 0.4$ in
plots b)-c), and detuning, $\Delta \le 0.2$ in plot d), but ruins at their
larger values, see plot e). Unlike \cite{Graefe_PRA_06}, we
have not found an improvement of the STIRAP transfer due to
the detuning $\Delta$. Instead, as is illustrated in the plot f), the
detuning mainly spoils the transport. As compared with \cite{Graefe_PRA_06},
we consider not one but three STIRAP steps. Nevertheless, even
in so long transfer chain, the transport of the interacting BEC
persists at $\Delta=0$. Altogether, Fig. 4 shows that STIRAP-like scheme
is quite reasonable for BEC transport in the triple-well trap \cite{Rab}.

\begin{figure}
\includegraphics[height=8.5cm,width=6.0cm,angle=-90]{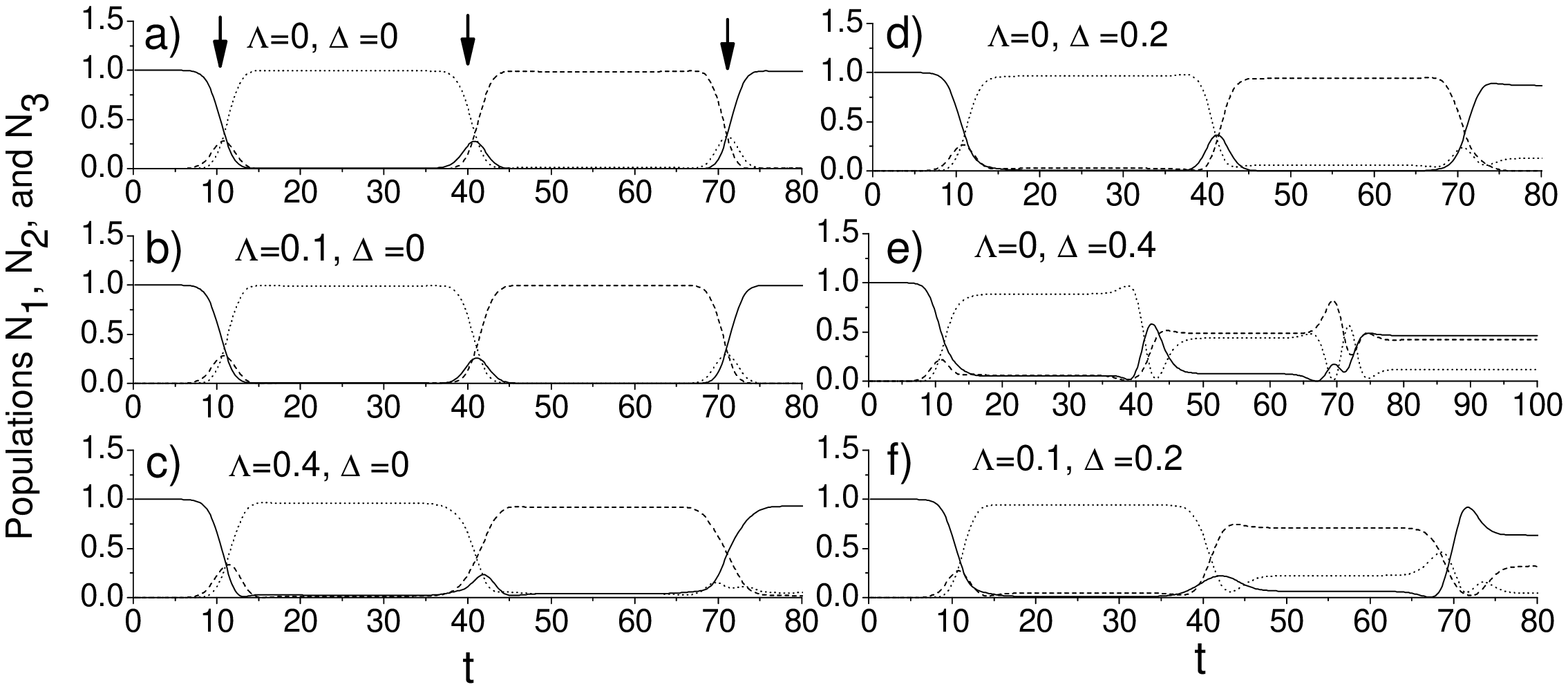}
\caption{\label{fig:fig6_3pop}
STIRAP transport of BEC in the circular well chain. The plots show
evolution of the populations $N_1$ (solid line),  $N_2$ (dash line) and
$N_3$ (dotted line) at different values of the interaction and detuning,
as indicated. The vertical arrows mark Stocks-pump coupling pairs. The
coupling width and shift are $\Gamma$=3.7 and $d$=2.
The time is dimensionless. The initial phase differences are zero.
}
\end{figure}

Note that transport in Fig. 4 is not fully
adiabatic. Even in the best cases we see small peaks signifying
a temporary slight populations of the intermediate states, hence
deviation from the strict adiabatic case \cite{Vitanov,Berg_STIRAP}.
In fact we have here quasiadiabatic transfer, though with a high
fidelity. In principle, one may improve the adiabaticity upgrading
the transfer parameters \cite{Ne_BEC}. However, this is
not the aim of the present paper. Instead, we are interested
in the effective and complete BEC transport between the initial and
target states, regardless of the temporary weak population of the
intermediate state. Moreover, our calculations show that the complete transfer
can be done even at intuitive sequence of the couplings when the pump
precedes Stokes \cite{Ne_BEC}, i.e. in strictly non-adiabatic case.
However, the adiabatic transfer has some advantages, e.g. it is less sensitive
to the parameters of the process.

In summary, we propose simple and effective adiabatic population
transfer schemes for the complete and irreversible transport of
BEC in double-and triple-well traps. The results are straightforwardly
generalized for multi-well traps with $M>3$. The schemes
work in the presence of a modest nonlinearity caused
by interatomic interaction. Our developments open
interesting perspectives for investigation of various geometric phases
\cite{Ne_BEC}, e.g. using the prescription \cite{Balac}.
There could be also various applications for Josephson qubits
\cite{JQ}, adiabatic quantum computing \cite{AQC},
topological states \cite{topol_1,topol_2}, etc.
At zero interaction, $\Lambda = 0$, our proposals are relevant
for the transport of individual atoms.

The work was supported by grants PVE 0067-11/2005 (CAPES, Brazil)
and 08-0200118 (RFBR, Russia).


\begin{thebibliography}{50}
\bibitem{note}
The basis is formed from the oscillator ground state
wave functions with shifted coordinates,
$\Phi_k ({\vec r})= u_({\vec r}-{\vec r}_k)$,
where ${\vec r}_k$ marks the minimum of the isolated k-th oscillator well.
For separated wells we have  $\int d\vec r \Phi^*_k\Phi_j = \delta_{kj}$.
\end{thebibliography}
\end{document}